\documentclass[conference]{IEEEtran}
\IEEEoverridecommandlockouts
\usepackage{cite}
\usepackage{amsmath,amssymb,amsfonts}
\usepackage{graphicx}
\usepackage{textcomp}
\usepackage{url}
\usepackage{hyperref}
\usepackage{xcolor}
\usepackage{tikz}
\usepackage{algorithm}
\usepackage{algorithmicx}
\usepackage{algpseudocode}
\def\BibTeX{{\rm B\kern-.05em{\sc i\kern-.025em b}\kern-.08em
    T\kern-.1667em\lower.7ex\hbox{E}\kern-.125emX}}
\begin{document}

\algdef{SE}[SUBALG]{Indent}{EndIndent}{}{\algorithmicend\ }%
\algtext*{Indent}
\algtext*{EndIndent}

\renewcommand{\algorithmicrequire}{\textbf{Input:}}
\renewcommand{\algorithmicensure}{\textbf{Output:}}

\title{Faster arbitrary-precision dot product and matrix multiplication
}


\author{\IEEEauthorblockN{Fredrik Johansson}
\IEEEauthorblockA{LFANT -- Inria Bordeaux\\
Talence, France \\
fredrik.johansson@gmail.com}
}

\maketitle

\begin{abstract}
We present algorithms for real and complex dot product and
matrix multiplication in
arbitrary-precision floating-point and ball arithmetic.
A low-overhead dot product is implemented on the level
of GMP limb arrays; it is about twice as fast as previous code in
MPFR and Arb at precision up to several hundred bits.
Up to 128 bits, it is 3-4 times as fast, costing
20-30 cycles per term for floating-point evaluation
and 40-50 cycles per term for balls.
We handle large matrix multiplications even
more efficiently
via blocks of scaled integer matrices.
The new methods are implemented in Arb and significantly
speed up polynomial operations and
linear algebra.
\end{abstract}


\section{Introduction}
The dot product and matrix multiplication are
core building blocks
for many numerical algorithms.
Our goal is to optimize these operations
in real and complex arbitrary-precision arithmetic.
We treat both floating-point arithmetic and
ball arithmetic~\cite{vdH:ball} in which errors are tracked rigorously using
midpoint-radius $[m \pm r]$ intervals.
Our implementations are part of the
open source (LGPL) Arb library~\cite{Johansson2017arb} (\url{http://arblib.org/}) as of version 2.16.

In this work, we only consider CPU-based software arithmetic
using GMP~\cite{granlund2017} for low-level operations
on mantissas represented by
arrays of 32-bit or 64-bit words (limbs).
This is the format used in MPFR~\cite{fousse2007mpfr} as well as Arb.
The benefit is flexibility (we handle mixed precision from
few bits to billions of bits);
the drawback is high bookkeeping overhead and limited vectorization opportunities.
In ``medium'' precision (up to several hundred bits),
arithmetic based on floating-point vectors (such as double-double and quad-double arithmetic)
offers higher performance on modern hardware
with wide SIMD floating-point units (inluding GPUs)~\cite{hida2007library,van2015faster,joldes2016arithmetic}.
However, such formats typically give up some flexibility (having a limited exponent range, usually assuming a fixed precision for all data).

The MPFR developers recently optimized
arithmetic for same-precision operands
up to 191 bits \cite{lefevre2017optimized}.
In this work, we reach even higher speed
without restrictions on the operands
by treating a whole dot product
as an atomic operation.
This directly speeds up many ``basecase'' algorithms
expressible using dot products, such as classical $O(N^2)$ polynomial multiplication
and division
and $O(N^3)$ matrix multiplication.
Section~\ref{sect:dot} describes the new dot product algorithm in detail.

For large polynomials and matrices (say, of size $N > 50$),
reductions to fast polynomial and matrix multiplication are
ultimately more efficient than iterated dot products.
Section~\ref{sect:matmul} looks at fast and accurate
matrix multiplication via scaled integer matrices.
Section \ref{sect:benchmarks} presents benchmark results
and discusses the combination of small-$N$ and large-$N$ algorithms
for polynomial operations and linear algebra.

\section{Precision and accuracy goals}

Throughout this text, $p \ge 2$ denotes the output target precision in bits.
For a dot product $\smash{s = \sum_{i=0}^{N-1} x_i y_i}$ where $x_i, y_i$
are floating-point numbers (not required to have the same precision), we aim to approximate $s$
with error of order $\smash{\varepsilon \sim 2^{-p} \sum_{i=0}^{N-1} |x_i y_i|}$.
In a practical sense, this accuracy is nearly optimal in $p$-bit arithmetic; up to cancellations that are unlikely
for generic data, uncertainty in the input will typically exceed~$\varepsilon$.
Since~$p$ is arbitrary,
we can set it to (say) twice the input precision
for specific tasks such as residual calculations.
To guarantee an error of $\smash{2^{-p} s}$ or even $p$-bit correct rounding of~$s$,
we may do a fast calculation as above with (say) $p+20$ bits of precision
and fall back to a slower correctly rounded summation~\cite{lefevre2017correctly}
only when the fast dot product fails.

A dot product in ball arithmetic becomes
\begin{equation*}
\textstyle{\sum_{i=0}^{N-1} [m_i \pm r_i] [m'_i \pm r'_i] \subseteq [m \pm r],}
\end{equation*}
\begin{equation*}
\textstyle{m\!=\!\sum_{i=0}^{N-1}\!m_i m'_i + \varepsilon, \, r \ge |\varepsilon| + \sum_{i=0}^{N-1}\!|m_i| r'_i \!+\! |m'_i| r_i \!+\! r_i r'_i}. \\
\end{equation*}
We compute $m$ with $p$-bit precision (resulting in some rounding error $\varepsilon$),
and we compute a low-precision upper bound for~$r$
that is tight up to rounding errors on $r$ itself.
If the input radii $r_i, r_i'$ are all zero and the computation of $m$ is exact ($\varepsilon = 0$),
then the output radius $r$ will be zero.
If $r$ is large, we can sometimes automatically reduce the precision
without affecting the accuracy of the output ball.



We require that matrix multiplication give
each output entry with optimal (\emph{up to cancellation})
accuracy, like the classical algorithm of evaluating $N^2$ separate dot products.
In particular, for a structured or badly scaled ball matrix like
\[
\begin{small}
\begin{pmatrix}
[1.23 \!\cdot\! 10^{100} \pm 10^{80}]\!\! & -1.5 & 0 \\
1 & \!\![2.34 \pm 10^{-20}]\! & [3.45 \pm 10^{-50}] \\
0 & 2 & \!\![4.56 \cdot 10^{-100} \pm 10^{-130}]
\end{pmatrix}\!,
\end{small}
\]
we preserve small entries and the individual error magnitudes.
Many techniques for fast multiplication
sacrifice such information.
Losing information is sometimes the right tradeoff,
but can lead to disaster (for example, 100-fold slowdown~\cite{Johansson2013arb})
when the input data is expensive to compute to high accuracy.
Performance concerns aside, preserving entrywise information
reduces the risk of surprises for users.

\section{Arbitrary-precision dot product}

\label{sect:dot}

The obvious algorithm to evaluate a dot product $\sum_{i=0}^{N-1} x_i y_i$
performs one multiplication followed by $N-1$ multiplications
and additions (or fused multiply-add operations) in a loop.
The functions
\texttt{arb\_dot}, \texttt{arb\_approx\_dot}, \texttt{acb\_dot}
and \texttt{acb\_approx\_dot} were introduced
in Arb 2.15 to replace most such loops. The function

\begin{footnotesize}
\begin{verbatim}
void arb_dot(arb_t res, const arb_t initial, int sub,
    arb_srcptr x, long xstep, arb_srcptr y,
    long ystep, long N, long p)
\end{verbatim}
\end{footnotesize}
sets \emph{res} to a ball containing
$\textstyle{\textit{initial} + (-1)^{\textit{sub}} \sum_{i=0}^{N-1} x_i \, y_i}$
where $x_i = [m_i \pm r_i]$, $y_i = [m_i' \pm r_i']$ and $\textit{initial}$ are balls of type \texttt{arb\_t}
given in arrays with strides of \emph{xstep} and \emph{ystep}.
The optional $\textit{initial}$ term (which may be \texttt{NULL}),
\emph{sub} flag and pointer stride lengths (which may be negative)
permit expressing many common operations
in terms of \texttt{arb\_dot} without extra arithmetic operations or
data rearrangement.

The \emph{approx} version is similar but ignores
the radii and computes
an ordinary floating-point dot product over the midpoints,
omitting error bound calculations.
The \emph{acb} versions are the counterparts for complex numbers.
All four functions are based on Algorithm~\ref{alg:dot}, which is explained in
detail below.

\subsection{Representation of floating-point numbers}

The dot product algorithm is designed around the
representation of arbitrary-precision
floating-point numbers and midpoint-radius intervals (balls) used in Arb.
In the following, $\beta \in \{32, 64\}$ is the word (limb) size,
and $p_{\operatorname{rad}} = 30$ is the
radius precision, which is a constant.

An \texttt{arb\_t} contains a midpoint of type \texttt{arf\_t}
and a radius of type \texttt{mag\_t}.
An \texttt{arf\_t} holds one of the special
values $0, \pm \infty, \operatorname{NaN}$, or
a regular floating-point value
\begin{equation}
\textstyle{m = (-1)^{\textit{sign}} \cdot 2^e \cdot \sum_{k=0}^{n-1} b_k 2^{\beta (k-n)}}
\label{eq:arf}
\end{equation}
where $b_k$ are $\beta$-bit mantissa limbs normalized
so that $2^{\beta-1} \le b_{n-1} \le 2^{\beta}-1$ and $b_0 \ne 0$.
Thus $n$ is always the minimal number of limbs needed to represent $x$,
and we have $2^{e-1} \le |m| < 2^e$.
The limbs are stored inline in the \texttt{arf\_t}
structure when $n \le 2$ and otherwise in heap-allocated memory.
The exponent $e$ can be arbitrarily large: a single word
stores $|e| < 2^{\beta-2}$ inline and larger $e$ as
a pointer to a GMP integer.

A \texttt{mag\_t} holds an unsigned floating-point value
$0, +\infty$, or
$r = (b / 2^{p_{\operatorname{rad}}}) 2^f$ where
$2^{p_{\operatorname{rad}}-1} \le b < 2^{p_{\operatorname{rad}}}$
occupies the low $p_{\operatorname{rad}}$ bits of one word.
We have $2^{f-1} \le |r| < 2^f$, and as for \texttt{arf\_t},
the exponent $f$ can be arbitrarily large.

The methods below
can be adapted for MPFR with minimal changes.
MPFR variables (\texttt{mpfr\_t}) use the
same representation as \eqref{eq:arf} except that a precision
$p$ is stored in the variable, the number of limbs is always
$n = \lceil p / \beta \rceil$ even if $b_0 = 0$,
there is no $n \le 2$ allocation optimization,
the exponent $e$ cannot be arbitrarily large,
and $-0$ is distinct from $+0$.

\subsection{Outline of the dot product}

We describe the algorithm for $\emph{initial} = 0$ and
$\emph{sub} = 0$. The general case can be viewed as extending
the dot product to length $N + 1$,
with trivial sign adjustments.

The main observation is that each arithmetic operation
on floating-point numbers of the form \eqref{eq:arf} has a lot of overhead
for limb manipulations (case distinctions, shifts, masks),
particularly during additions and subtractions.
The remedy is to use a fixed-point accumulator for the whole
dot product and only convert to a rounded and normalized floating-point number
at the end. The case distinctions for subtractions
are simplified by using two's complement arithmetic.
Similarly, we use a fixed-point accumulator
for the radius dot product.

We make two passes over the data: the first pass inspects all terms,
looks for exceptional cases, and determines an appropriate working
precision and exponents to use for the accumulators.
The second pass evaluates the dot product.

There are three sources of error: arithmetic error on the accumulator
(tracked with one limb counting ulps), the final rounding error,
and the propagated error from the input balls.
At the end, the three contributions are added to a single $p_{\operatorname{rad}}$-bit
floating-point number. The \emph{approx} version of the
dot product simply omits all these steps.

Except where otherwise noted, all quantities
describing exponents, shift counts (etc.) are single-word ($\beta$-bit) signed integers
between $\textit{MIN} = -2^{\beta-1}$ and $\textit{MAX} = 2^{\beta-1}-1$,
and all limbs are $\beta$-bit unsigned words.

\begin{algorithm}[h]
  \caption{Dot product in arbitrary-precision ball arithmetic: given $[m_i \pm r_i], [m'_i \pm r'_i], 0 \le i < N$ and a precision $p \ge 2$, compute $[m \pm r]$ containing $\sum_{i=0}^{N-1} [m_i \pm r_i]  [m'_i \pm r'_i]$.}
  \label{alg:dot}
  \begin{algorithmic}[1]
    \State \textbf{Setup:} check unlikely cases (infinities, NaNs, overflow and underflow); determine exponent $e_s$ and number of limbs $n_s$ for the midpoint accumulator; determine exponent $e_{\operatorname{rad}}$ for the radius accumulator; reduce $p$ if possible. (See details below.)
  \State \textbf{Initialization:} allocate temporary space; initialize accumulator limbs $s$: $s_{n_s-1}, \ldots, s_0 \gets 0, \ldots, 0$, one limb $\textit{err} \gets 0$ for the ulp error on $s_0$, and a 64-bit integer $\textit{srad} \gets 0$ as radius accumulator.
  \State \textbf{Evaluation:} for each term $t = m_i m'_i$, compute the limbs of $t$ that overlap with $s$, shift and add to $s$ (or two's complement subtract if $t < 0$), incrementing $\textit{err}$ if inexact. Add scaled upper bound for $|m_i| r'_i + |m'_i| r_i + r_i r_i'$ to $\textit{srad}$.
  \State \textbf{Finalization:}
  \begin{enumerate}
    \item If $s_{n_s-1} \ge 2^{\beta-1}$, negate $s$ (one call to GMP's \texttt{mpn\_neg}) and set $\textit{sign} \gets 1$, else set $\textit{sign} \gets 0$.
    \item $m \gets (-1)^{\textit{sign}} \cdot 2^{e_s} \cdot (\sum_{k=0}^{n_s-1} s_k 2^{\beta (k-n_s)})$ rounded to $p$ bits, giving a possible rounding error $\varepsilon_{\text{round}}$.
    \item $r \gets \varepsilon_{\text{round}} + \textit{err} \cdot  2^{e_s - n_s \beta} + \textit{srad} \cdot 2^{e_{\operatorname{rad}} - p_{\operatorname{rad}}}$ as a floating-point number with $p_{\operatorname{rad}}$ bits (rounded up).
    \item Free temporary space and output $[m \pm r]$.
  \end{enumerate}
  \end{algorithmic}
\end{algorithm}

\subsection{Setup pass}
\label{sect:setuppass}

The setup pass in Algorithm~\ref{alg:dot} uses the following steps:

\begin{enumerate}
\item $N_{\operatorname{nonzero}} \gets 0$ (number of nonzero terms).
\item $e_{\operatorname{max}} \gets \textit{MIN}$ (upper bound for term exponents).
\item $e_{\operatorname{min}} \gets \textit{MAX}$ (lower bound for content).
\item $e_{\operatorname{rad}} \gets \textit{MIN}$ (upper bound for radius exponents).
\item For $0 \le i < N$:
\begin{enumerate}
\item If any of $m_i, m_i', r_i, r_i'$ is non-finite or has an exponent outside of $\pm 2^{\beta-4}$ (unlikely), quit Algorithm~\ref{alg:dot} and use a fallback method.
\item If $m_i$ and $m_i'$ are both nonzero, with respective exponents $e, e'$ and limb counts $n, n'$:
\begin{itemize}
\item Set $N_{\operatorname{nonzero}} \gets N_{\operatorname{nonzero}} + 1$.
\item Set $e_{\operatorname{max}} \gets \max(e_{\operatorname{max}}, \, e + e')$.
\item If $p > 2 \beta$, $e_{\operatorname{min}} \gets \min(e_{\operatorname{min}}, \, e + e' - \beta (n + n'))$.
\end{itemize}
\item For each product $|m_i| r_i'$, $|m_i'| r_i$, $r_i r_i'$ that is nonzero,
denote the exponents of the respective factors by $e, e'$ and set $e_{\operatorname{rad}} \gets \max(e_{\operatorname{rad}}, e + e')$.
\end{enumerate}
\item If $e_{\operatorname{max}}\!=\!e_{\operatorname{rad}}\!=\!\textit{MIN}$, quit Algorithm~\ref{alg:dot} and output $[0 \pm 0]$
\item (Optimize $p$.) If $e_{\operatorname{max}} = \textit{MIN}$, set $p \gets 2$. Otherwise:
\begin{enumerate}
\item If $e_{\operatorname{rad}} \ne \textit{MIN}$, set $p \gets \min(p, e_{\operatorname{max}} - e_{\operatorname{rad}} + p_{\operatorname{rad}})$ (if the final radius $r$ will be larger than the expected arithmetic error, we can reduce the precision used to compute $m$ without affecting the accuracy of the ball $[m \pm r]$).
\item If $e_{\operatorname{min}} \ne \textit{MAX}$, set $p \gets \min(p, e_{\operatorname{max}} - e_{\operatorname{min}} + p_{\operatorname{rad}})$ (if all terms fit in a window smaller than $p$ bits, reducing the precision does not change the result).
\item Set $p \gets \max(p, 2)$.
\end{enumerate}
\item Set $\textit{padding} \gets 4 + \operatorname{bc}(N)$, where $\operatorname{bc}(\nu) = \lceil \log_2(\nu+1) \rceil$ denotes the binary length of $\nu$.
\item Set $\textit{extend} \gets \operatorname{bc}(N_{\operatorname{nonzero}}) + 1$.
\item Set $n_s \gets \max(2, \lceil (p + \textit{extend} + \textit{padding}) / \beta \rceil)$.
\item Set $e_s \gets e_{\operatorname{max}} + \textit{extend}$.
\end{enumerate}

All terms $|m_i m_i'|$ are bounded by $2^{e_{\operatorname{max}}}$ and similarly all radius
terms are bounded by $2^{e_{\operatorname{rad}}}$.
The width of the accumulator is $p$ bits plus $\textit{extend}$ leading bits and $\textit{padding}$ trailing bits, rounded up to a whole number of limbs $n_s$.
The quantity $\textit{extend}$ guarantees that carries never overflow the leading limb $s_{n_s-1}$, including one bit for two's complement negation;
it is required to guarantee correctness.
The quantity $\textit{padding}$ adds a few guard bits to enhance the accuracy of the dot product; this is an entirely optional tuning parameter.

\subsection{Evaluation}

For a midpoint term $m_i m_i' \ne 0$, denote the exponents of $m_i, m_i'$ by $e, e'$ and
the limb counts by $n, n'$. The multiply-add
operation uses the following steps.

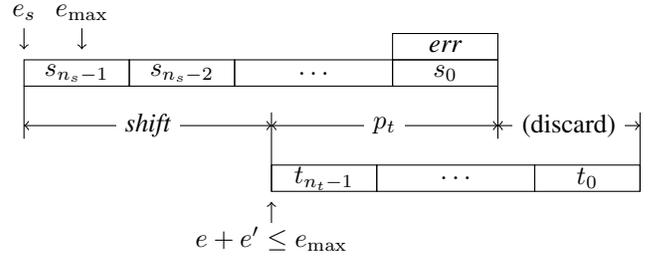
\begin{figure}
\begin{center}
\begin{tikzpicture}[scale=0.7]

\node[align=right] at (0,1.4) {$e_s$};
\draw[->] (0,1.1) -- (0,0.7);

\node[align=right] at (1.1,1.4) {$e_{\operatorname{max}}$};
\draw[->] (1.1,1.1) -- (1.1,0.7);

\draw [draw=black] (0,0) rectangle (2,0.5);
\node[align=center] at (1,0.25) {$s_{n_s-1}$};
\draw [draw=black] (2,0) rectangle (4,0.5);
\node[align=center] at (3,0.25) {$s_{n_s-2}$};
\draw [draw=black] (4,0) rectangle (7,0.5);
\node[align=center] at (5.5,0.25) {$\ldots$};
\draw [draw=black] (7,0) rectangle (9,0.5);
\node[align=center] at (8,0.25) {$s_0$};
\draw [draw=black] (7,0.5) rectangle (9,1);
\node[align=center] at (8,0.75) {$\textit{err}$};

\draw [draw=black] (4.7,-1 -1) rectangle (6.7,-0.5 -1);
\node[align=center] at (5.7,-0.75 -1) {$t_{n_t-1}$};
\draw [draw=black] (6.7,-1 -1) rectangle (9.7,-0.5 -1);
\node[align=center] at (6.7+1.5,-0.75 -1) {$\ldots$};
\draw [draw=black] (9.7,-1 -1) rectangle (11.7,-0.5 -1);
\node[align=center] at (10.7,-0.75 -1) {$t_0$};

\draw (0,0) -- (0,-1);
\draw[<->] (0,-0.75) -- (4.7,-0.75);
\draw (4.7,-0.5) -- (4.7,-2);
\draw[<->] (4.7,-0.75) -- (9,-0.75);
\draw (9,0) -- (9,-1);
\draw[<->] (9,-0.75) -- (11.7,-0.75);
\draw (11.7,-0.5) -- (11.7,-2);
\node[align=center,fill=white] at (2.35,-0.75) {$\textit{shift}$};
\node[align=center,fill=white] at (6.85,-0.75) {$p_t$};
\node[align=center,fill=white] at (10.35,-0.75) {(discard)};

\node[align=right] at (4.7,-2.9) {$e+e' \le e_{\operatorname{max}}$};
\draw[->] (4.7,-2.6) -- (4.7,-2.2);


\end{tikzpicture}
\end{center}

\caption{The accumulator $s_{n_s-1}, \ldots, s_0$ and the term $t_{n_t-1}, \ldots, t_0$, prior to limb alignment. More significant limbs are shown towards the left.}
\label{fig:accumulator}
\end{figure}

\begin{enumerate}
\item Set $\textit{shift} \gets e_s - (e + e')$, $\textit{shift\_bits} \gets \textit{shift} \bmod \beta$, $\textit{shift\_limbs} \gets \lfloor \textit{shift} / \beta \rfloor$.
\item If $\textit{shift} \ge \beta n_s$, set $\textit{err} \gets \textit{err} + 1$ and go on to the next term (this term does not overlap with the limbs in $s$).
\item Set $p_t \gets \beta n_s - \textit{shift}$ (effective bit precision needed for this term), and set $n'' \gets \lceil p_t / \beta \rceil + 1$. If $n > n''$ or $n' > n''$, set $\textit{err} \gets \textit{err} + 1$.
      (We read at most $n''$ leading limbs from $m$ and $m'$ since the smaller limbs have a negligible contribution to the dot product; in case of truncation, we increment the error bound by 1 ulp.)
\item Set $n_t \gets \min(n, n'') + \min(n', n'')$. The term will be stored in up to $n_t + 1$ temporary limbs $t_{n_t}, \ldots, t_0$ pre-allocated in the initialization of Algorithm~\ref{alg:dot}.
\item Set $t_{n_t-1}, \ldots, t_0$ to the product of the top $\min(n, n'')$ limbs of $m$ and the top $\min(n', n'')$ limbs of $m'$ (this is one call to GMP's \texttt{mpn\_mul}). We now have the situation depicted in Figure~\ref{fig:accumulator}.
\item (Bit-align the limbs.) If $\textit{shift\_bits} \ne 0$, set $t_{n_t}, \ldots, t_0$ to $t_{n_t-1}, \ldots, t_0$ right-shifted by $\textit{shift\_bits}$ bits (this is a pointer adjustment and one call to \texttt{mpn\_rshift}) and then set $n_t \gets n_t + 1$.
\item (Strip trailing zero limbs.) While $t_0 = 0$, increment the pointer to $t$ and set $n_t \gets n_t - 1$.
\end{enumerate}

It remains to add the aligned limbs of $t$ to the accumulator $s$.
We have two cases, with $v$ denoting the number of overlapping limbs
between $s$ and $t$ and $d_s$ and $d_t$ denoting the offsets
from $s_0$ and $t_0$ to the overlapping segment.
If $\textit{shift\_limbs} + n_t \le n_s$ (no discarded limbs), set $d_s \gets n_s - \textit{shift\_limbs} - n_t$, $d_t \gets 0$
and $v \gets n_t$.
Otherwise, set $d_s \gets 0$, $d_t \gets n_t - n_s + \textit{shift\_limbs}$, $v \gets n_s - \textit{shift\_limbs}$
and $\textit{err} \gets \textit{err} + 1$.
The addition is now done using the GMP code

\begin{footnotesize}
\begin{verbatim}
cy = mpn_add_n(s + ds, s + ds, t + dt, v);
mpn_add_1(s + ds + v, s + ds + v, shift_limbs, cy);
\end{verbatim}
\end{footnotesize}
if $m_i m_i' > 0$, or in case $m_i m_i' < 0$ using
\texttt{mpn\_sub\_n} and \texttt{mpn\_sub\_1}
to perform a two's complement subtraction.

Our code has two more optimizations.
If $n \le 2, n' \le 2$, $n_s \le 3$,
the limb operations are done using inline code
instead of calling GMP functions, speeding up
precision $p \le 128$ (on 64-bit machines).
When $p_t \ge 25\beta$ and $\min(n,n') \beta > 0.9 p_t$,
we compute $n''$ leading limbs of the product using the
MPFR-internal function \texttt{mpfr\_mulhigh\_n}
instead of \texttt{mpn\_mul}.
This is done with up to~1 ulp error on $s_0$ and is therefore accompanied by an extra increment of $\textit{err}$.

\subsection{Radius operations}

For the radius dot product
$\sum_{i=0}^{N-1} |m_i| r'_i + |m'_i| r_i + r_i r_i'$,
we convert the midpoints $|m_i|$, $|m'_i|$ to upper bounds in the radius format
$r = (b / 2^{p_{\operatorname{rad}}}) 2^e$
by taking the top $p_{\operatorname{rad}}$ bits of the top limb and
incrementing;
this results in the weakly normalized mantissa
$2^{p_{\operatorname{rad}}-1} \le b \le 2^{p_{\operatorname{rad}}}$.
The summation is done with
an accumulator $(\textit{srad} / 2^{p_{\operatorname{rad}}}) 2^{e_{\operatorname{rad}}}$ where $\textit{srad}$ is one unsigned 64-bit integer (1 or 2 limbs).
The step to add an upper bound for $(a / 2^{p_{\operatorname{rad}}}) (b / 2^{p_{\operatorname{rad}}}) 2^e$
is $\textit{srad} \gets \textit{srad} + \lfloor (a b) / 2^{p_{\operatorname{rad}} + e_{\operatorname{rad}} - e} \rfloor + 1$
if $e_{\operatorname{rad}} - e < p_{\operatorname{rad}}$ and $\textit{srad} \gets \textit{srad} + 1$ otherwise.

By construction, $e_{\operatorname{rad}} \ge e$, and due to the 34 free bits for carry accumulation,
$\textit{srad}$ cannot overflow if $N < 2^{32}$.
(Larger $N$ could be supported by increasing $e_{\operatorname{rad}}$,
at the cost of some loss of accuracy.)
We use conditionals to skip zero terms; the radius
dot product is therefore evaluated as zero whenever possible,
and if the input balls are exact, no radius computations are done
apart from inspecting the terms.

\subsection{Complex numbers}

Arb uses rectangular ``balls'' $[a \pm r] + [b \pm s] i$
to represent complex numbers.
A complex dot product is essentially performed as two
length-$2N$ real dot products.
This preserves information about whether
real or imaginary parts are exact or zero, and both parts can be
computed with high relative accuracy when they have different
scales.
The algorithm could be adapted
in the obvious way for true complex balls (disks).

For terms where both real and imaginary parts have similar
magnitude and high precision,
we use the additional optimization of avoiding one real multiplication
via the formula
\begin{equation}
(a+bi)(c+di) = ac - bd + i [(a+b)(c+d) - ac - bd].
\label{eq:threeprod}
\end{equation}
Since this formula is applied exactly and only for the midpoints,
accuracy is not compromised.
The cutoff occurs at the rather high 128 limbs (8192 bits)
since \eqref{eq:threeprod} is implemented using exact products
and therefore competes against mulhigh;
an improvement is possible
by combining mulhigh with \eqref{eq:threeprod}.

\section{Matrix multiplication}

\label{sect:matmul}

We consider the problem of multiplying an $M \times N$ ball matrix $[A \pm R_A]$ by an $N \times K$ ball matrix
$[B \pm R_B]$ (where $R_A, R_B$ are nonnegative matrices and $[\pm]$ is interpreted entrywise).
The \emph{classical algorithm} can be viewed as computing
$MP$ dot products of length $N$.
For large matrices, it is better to convert from arbitrary-precision
floating-point numbers to integers~\cite{vdH:ball}.
Integer matrices can be multiplied efficiently using multimodular
techniques, working modulo several word-size primes followed by
Chinese remainder theorem reconstruction.
This saves time
since computations done over a fixed word size have less overhead than arbitrary-precision computations.
Moreover, for modest $p$, the running time essentially scales as $O(p)$ compared to the $O(p^2)$
with dot products, as long as the cost of the modular reductions and reconstructions does not dominate.
The downside of converting floating-point numbers to integers is that we either must truncate entries (losing accuracy) or
zero-pad (losing speed).

Our approach to matrix multiplication
resembles methods for fast and accurate polynomial multiplication
discussed
in previous work~\cite{vdH:stablemult},\cite{Johansson2017arb}.
For polynomial multiplication, Arb scales the inputs and converts the coefficients to integers,
adaptively splitting the polynomials into smaller blocks
to keep the height of the integers small.
The integer polynomials are then multiplied using
FLINT~\cite{hart2010fast}, which selects between classical, Karatsuba and
Kronecker algorithms and an asymptotically fast
Sch\"{o}nhage-Strassen FFT.
Arb implements other operations (such as division)
via methods such as Newton iteration
that asymptotically reduce to polynomial multiplication.

In this section, we describe an approach to multiply matrices
in Arb
following similar principles.
We compute $[A \pm R_A][B \pm R_B]$ using
three products $AB$, $|A|R_B$, $R_A(|B|+R_B)$ where we use
FLINT integer matrices for the high-precision midpoint product $AB$.
FLINT in turn uses classical multiplication, the Strassen algorithm,
a multimodular algorithm employing 60-bit primes, and combinations
of these methods.
An important observation for both polynomials and matrices is that fast
algorithms such as Karatsuba, FFT and Strassen multiplication
do not affect accuracy when used on the integer level.

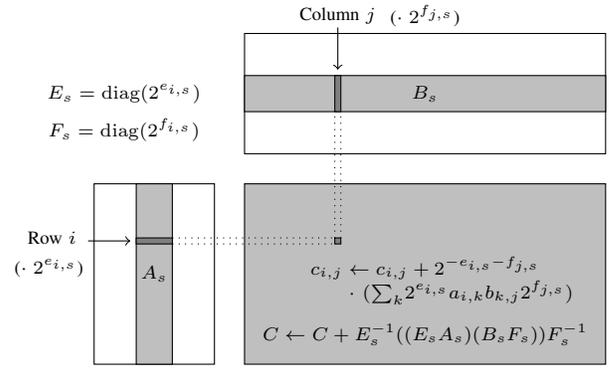
\begin{figure}
\begin{center}
\begin{tikzpicture}[scale=0.8]
\tikzset{font=\scriptsize}

\draw [draw=black] (0,0) rectangle (2,3);
\draw [draw=black] (2.5,3.5) rectangle (8.5,5.5);
\draw [draw=black,fill=lightgray] (2.5,0) rectangle (8.5,3.0);

\draw [draw=black,fill=lightgray] (0.7,0) rectangle (1.3,3);
\draw [draw=black,fill=lightgray] (2.5,4.2) rectangle (8.5,4.8);

\draw [draw=black,fill=gray] (0.7,2) rectangle (1.3,2.1);
\draw [draw=black,fill=gray] (4,4.2) rectangle (4.1,4.8);
\draw [draw=black,fill=gray] (4,2) rectangle (4.1,2.1);

\draw[dotted] (1.3,2) -- (4,2);
\draw[dotted] (1.3,2.1) -- (4,2.1);
\draw[dotted] (4,2.1) -- (4,4.2);
\draw[dotted] (4.1,2.1) -- (4.1,4.2);

\node[align=center] at (5.5,1.6) {$c_{i,j} \gets c_{i,j} + 2^{-e_{i,s}-f_{j,s}}$};
\node[align=center] at (6.1,1.2) {$\cdot \,\, (\sum_k \!2^{e_{i,s}} a_{i,k} b_{k,j} 2^{f_{j,s}})$};
\node[align=center] at (5.5,0.5) {$C \gets C + E_s^{-1} ((E_s A_s) (B_s F_s)) F_s^{-1}$};
\node[align=center] at (1,1.5) {$A_s$};
\node[align=center] at (5.5,4.5) {$B_s$};

\node[align=center] at (-0.7,2.1) {Row $i$};
\node[align=center] at (-0.75,1.6) {($\cdot \; 2^{{e_{i,s}}}$)};

\draw[->] (4.05,5.6) -- (4.05,4.9);
\draw[->] (-0.1,2.05) -- (0.6,2.05);

\node[align=center] at (4.05,5.8) {Column $j$};
\node[align=center] at (5.5,5.8) {($\cdot \; 2^{{f_{j,s}}}$)};

\node[align=center] at (0.5,4.5) {$E_s = \operatorname{diag}(2^{e_{i,s}})$};
\node[align=center] at (0.5,3.9) {$F_s = \operatorname{diag}(2^{f_{i,s}})$};

\end{tikzpicture}
\end{center}

\caption{Matrix multiplication $C = AB$ using scaled blocks.}
\label{fig:matblocks}
\end{figure}

\subsection{Splitting and scaling}

The earlier work by van der Hoeven~\cite{vdH:ball} proposed
multiplying arbitrary-precision matrices via integers truncated
to $p$-bit height,
splitting size-$N$ matrices into $m^2$
blocks of size $N/m$, where the user selects $m$ to
balance speed and accuracy.
Algorithm~\ref{alg:block} improves on this idea by
using a fully automatic and adaptive splitting strategy
that guarantees near-optimal entrywise accuracy (like
the classical algorithm).

\begin{algorithm}
  \caption{Matrix multiplication using blocks: given ball matrices $[A \pm R_A]$, $[B \pm R_B]$ and a precision $p \ge 2$, compute $[C \pm R_C]$ containing $[A \pm R_A] [B \pm R_B]$}
  \label{alg:block}
  \begin{algorithmic}[1]
  \State $[C \pm R_C] \gets [0 \pm 0]$ \Comment{Initialize the zero matrix}
  \State $h \gets 1.25 \min(p, \max(p_A, p_B)) \!+\! 192$, where $p_M$ is the minimum floating-point precision needed to represent all entries of $M$ exactly \Comment{Height bound tuning parameter}
  \State $S \gets \{0,\ldots,N-1\}$ where $N$ is the inner dimension
  \While{$S \ne \{\}$}
    \State Extract $s \subseteq S$ such that $|E_s A_s| < 2^h$ and $|B_s F_s| < 2^h$
    \If {$\operatorname{size}(s) < 30$} \Comment{Basecase for short blocks}
      \State Extend $s$ to $\min(30, \operatorname{size}(S))$ indices
      \State $[C \pm R_C] \gets [C \pm R_C] + A_s B_s$ (using dot products)
    \Else
      \State $T \gets (E_s A_s) (B_s F_s)$ \Comment{Matrix product over $\mathbb{Z}$}
      \State $[C \pm R_C] \gets [C \pm R_C] + E_s^{-1} T F_s^{-1}$ \Comment{Ball addition, with possible rounding error on $C$ added to $R_C$}
    \EndIf
    \State $S \gets S \setminus s$
  \EndWhile
  \State Compute $R_1 = |A|R_B$ and $R_2 = R_A(|B|+R_B)$ by splitting and scaling into blocks of \texttt{double} as above (but using floating-point arithmetic with upper bounds instead of ball arithmetic), and set $R_C \gets R_C + R_1 + R_2$
  \State Output $[C \pm R_C]$
  \end{algorithmic}
\end{algorithm}

We split $A$ into column submatrices $A_s$ and $B$ into
row submatrices $B_s$, where $s$ is some subset of the indices.
For any such $A_s$, and for each row index $i$,
let $e_{i,s}$ denote the unique scaling exponent such that row~$i$ of
$2^{e_{i,s}} A_s$ consists of integers of minimal height
($e_{i,s}$ is uniquely determined unless row $i$ of $A_s$ is identically zero,
in which case we may take $e_{i,s} = 0$).
Similarly let $f_{j,s}$ be the optimal
scaling exponent for column $j$ of $B_s$.
Then the contribution of $A_s$ and $B_s$ to $C = AB$ consists of
$E_s^{-1} ((E_s A_s) (B_s F_s)) F_s^{-1}$
where $E_s = \operatorname{diag}(2^{e_{i,s}})$ scales the rows of
$A_s$ and $F_s = \operatorname{diag}(2^{f_{i,s}})$ scales the
columns of $B_s$ (see Figure~\ref{fig:matblocks}),
and where we may multiply $(E_s A_s) (B_s F_s)$
over the integers.

Crucially, only magnitude variations
\emph{within} rows of $A_s$ (columns of $B_s$) affect the height;
the rows of $A_s$ can have different
magnitude from each other (and similarly for $B_s$).

We extract indices $s$ by performing a greedy search in increasing order,
appending columns to $A_s$ and rows to $B_s$ as long as a
height bound is satisfied.
The tuning parameter $h$ balances the advantage of using larger blocks against the
disadvantage of using larger zero-padded integers.
In the common case where both $A$ and $B$ are uniformly scaled
and have the same (or smaller) precision as the output,
one block product is sufficient.
One optimization is omitted from the pseudocode:
we split the rectangular matrices $E_s A_s$ and $B_s F_s$
into roughly square blocks before carrying out the multiplications.
This reduces memory usage for the temporary
integer matrices and can reduce the heights of the blocks.

We compute the radius products $|A|R_B$ and $R_A(|B|+R_B)$ (where $|A|$ and $|B|$ are rounded to $p_{\textrm{rad}}$ bits)
via \texttt{double} matrices, using a similar block strategy.
The \texttt{double} type has a normal exponent range of
$-1022$ to $1023$, so if we set $h = 900$ and
center $E_s A_s$ and $B_s F_s$ on this range,
no overflow or underflow can occur.
In practice a single block is sufficient for most matrices arising in medium precision
computations.

\subsection{Improvements}

Algorithm~\ref{alg:block} turns out to perform reasonably well in practice
when many blocks are used, but it could certainly be improved.
The bound $h$ could be tuned, and the greedy strategy
to select blocks is clearly not always optimal.
Going even further, we could extract non-contiguous submatrices,
add an extra inner scaling matrix $G_s G_s^{-1}$
for the columns of $A_s$ and rows of $B_s$,
and combine scaling with permutations.
Finding the best strategy for
badly scaled or structured matrices appears to be a difficult problem.
There is some resemblance to the balancing
problem for eigenvalue algorithms~\cite{kressner2005}.

Both Algorithm~\ref{alg:block} and the analogous algorithm
used for polynomial multiplication in Arb have the disadvantage that all
input bits are used, unlike classical multiplication based on Algorithm~\ref{alg:dot}
which omits negligible limbs.
This important optimization for non-uniform polynomials (compare~\cite{vdH:stablemult}) and matrices
should be considered in future work.

\subsection{Complex matrices}

We multiply complex matrices using
four real matrix multiplications
$(A+Bi)(C+Di) = (AC - BD) + (AD + BC) i$ outside of the basecase
range for using complex dot products.
An improvement would be to use \eqref{eq:threeprod} to multiply
the midpoint matrices when all entries are uniformly scaled; \eqref{eq:threeprod}
could also be used for blocks
with a splitting and scaling strategy that considers the real and imaginary
parts simultaneously.

\section{Benchmarks}

\label{sect:benchmarks}

Except where noted, the following results were obtained on an Intel i5-4300U CPU
using GMP 6.1, MPFR 4.0, MPC 1.1~\cite{enge2018mpc}
(the complex extension of MPFR),
QD 2.3.22~\cite{hida2007library} (106-bit double-double and 212-bit quad-double arithmetic),
Arb 2.16, and the December 2018
git version of FLINT.

\subsection{Single dot products}

\begin{figure}[tbp]
\centerline{\includegraphics[width=0.95 \linewidth]{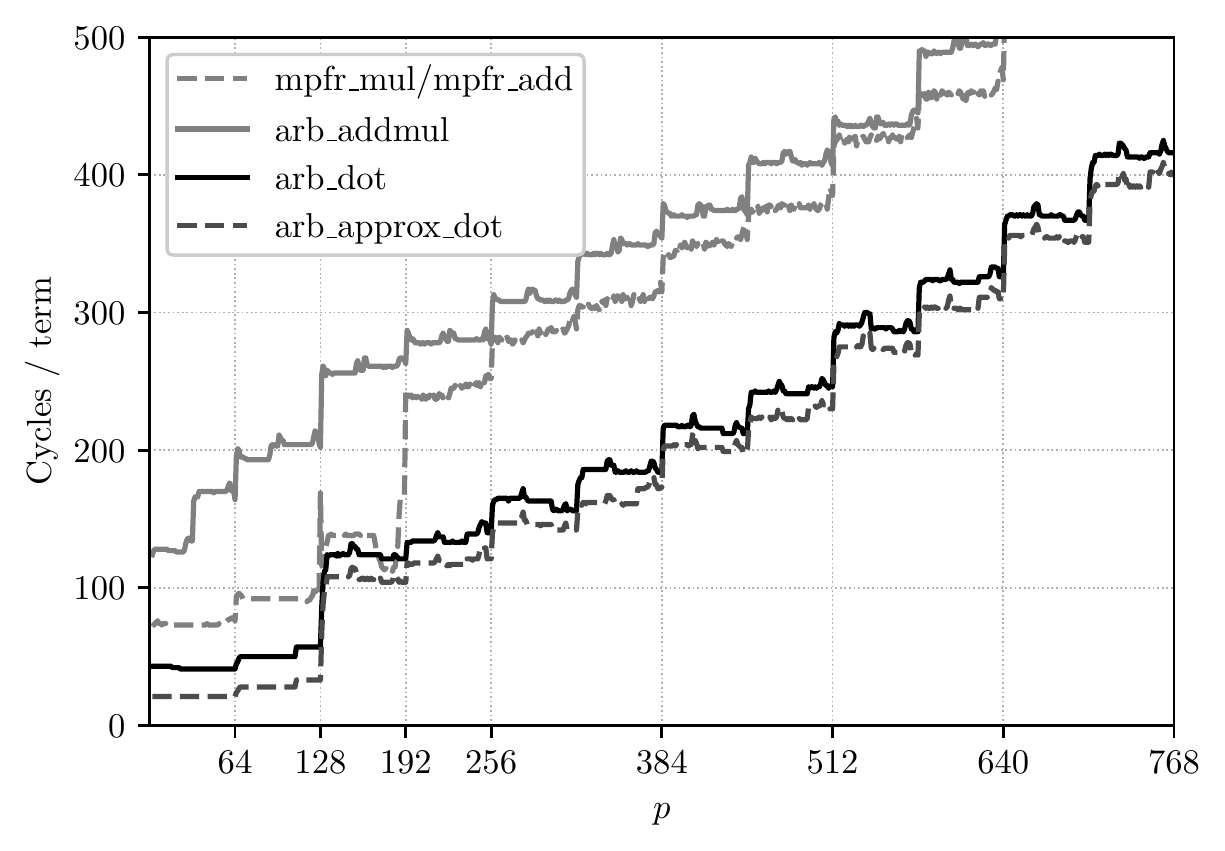}}
\vspace{-2mm}
\caption{Dot product cost (cycles/term) as a function of the precision $p$, using MPFR, simple Arb code (\texttt{arb\_addmul} in a loop), and Algorithm~\ref{alg:dot} in Arb (both ball and approximate floating-point versions).}
\vspace{-2mm}
\label{fig:plotdot}
\end{figure}

Figure~\ref{fig:plotdot} and Table~\ref{tab:dotcycles} show
timings measured in CPU cycles per term
for a dot product of length $N = 100$
with uniform $p$-bit entries.
We compare a simple loop using QD
arithmetic, three MPFR versions,
a simple Arb loop (\emph{addmul} denoting repeated
multiply-adds with \texttt{arb\_addmul}),
and Algorithm~\ref{alg:dot} in Arb, both for
balls (\emph{dot} denoting \texttt{arb\_dot})
and floating-point numbers (\emph{approx} denoting \texttt{arb\_approx\_dot}).
Similarly, we include results for complex dot products, comparing
MPC and three Arb methods.
The \emph{mul/add} MPFR version uses \texttt{mpfr\_mul}
and \texttt{mpfr\_add}, with a preallocated temporary variable;
\emph{fma} denotes multiply-adds with \texttt{mpfr\_fma};
our \emph{sum} code
writes exact products to an array and calls \texttt{mpfr\_sum}~\cite{lefevre2017correctly}
to compute the sum (and hence the dot product) with correct rounding.
We make several observations:

\begin{itemize}
\item The biggest improvement is seen for $p \le 128$ (up to two limbs). The ball dot product is up to 4.2 times faster than the simple Arb loop
(and 2.0 times faster than MPFR); the \emph{approx} version is up to 3.7 times faster than MPFR.
\item A factor 1.5 to 2.0 speedup persists up to several hundred bits, and the speed for very large $p$ is close to the optimal throughput for GMP-based multiplication.
\item Ball arithmetic error propagation adds 20 cycles/term overhead, equivalent to a factor 2.0 when $p \le 128$ and a negligible factor at higher precision.
\item At $p = 106$, the \emph{approx} dot product is about as fast as QD double-double arithmetic, while
the ball version is half as fast; at $p = 212$, either version is twice as fast as QD quad-double arithmetic.
\item Complex arithmetic costs quite precisely four times more than real arithmetic. The speedup of our code compared to MPC is even greater than compared to MPFR.
\item A future implementation of a correctly rounded dot product for MPFR and MPC using Algorithm~\ref{alg:dot} with \texttt{mpfr\_sum} as a fallback
should be able to achieve nearly the same average speed as the \emph{approx} Arb version.
\end{itemize}

\begin{table}
\caption{Cycles/term to evaluate a dot product ($N = 100$).}
\begin{center}
\renewcommand{\arraystretch}{1.05}
\begin{tabular}{|r | r | r r r | r r r|}
\hline
   & QD & \multicolumn{3}{|c|}{MPFR (real)} & \multicolumn{3}{|c|}{Arb (real)} \\
$p$ & & \!\!mul/add & fma & sum & \!\!addmul & dot & \!\!approx \\ \hline
53 &   & 74 & 99 & 108 & 169 & 40 & 20 \\
106 & 26 &  97 & 156 & 124 & 203 & 49 & 27 \\
159 & &  140 & 183 & 169 & 257 & 123 & 105 \\
212 & 265 &  237 & 208 & 188 & 277 & 133 & 117 \\
424 & &  350 & 288 & 288 & 374 & 215 & 201 \\
848 & &  670 & 619 & 597 & 705 & 522 & 499 \\
1696 & &   1435 & 1675 & 1667 & 1823 & 1471 & 1451 \\
3392 & &  4059 & 4800 & 4741 & 4875 & 3906 & 3880 \\
\!\!13568 & &  \!33529 & \!39546 & \!39401 & \!39275 & \!32476 & \!32467 \\
\hline
   &   & \multicolumn{3}{|c|}{MPC (complex)} & \multicolumn{3}{|c|}{Arb (complex)} \\
$p$ & & \!\!mul/add & & & \!\!addmul & dot & \!\!approx \\ \hline
53 & & 570 & & & 772 & 166 & 84 \\
106 & & 885 & & & 911 & 208 & 112 \\
159 & & 1016 & & & 1243 & 499 & 419 \\
212 & & 1123 & & & 1346 & 555 & 478 \\
424 & & 1591 & & & 1735 & 882 & 775 \\
848 & & 2803 & & & 3054 & 2097 & 2045 \\
1696 & & 8355 & & & 6953 & 5889 & 5821 \\
3392 & & \!18527 & & & \!17926 & \!15691 & \!15618 \\
\!\!13568 & & \!\!129293 & & & \!\!127672 & \!\!125757 & \!\!125634 \\
\hline
\end{tabular}
\label{tab:dotcycles}
\end{center}
\vspace{-2mm}
\end{table}

\begin{table}
\caption{Cycles/term to evaluate a dot product, variable $N$.}
\begin{center}
\renewcommand{\arraystretch}{1.05}
\begin{tabular}{|r | r r r r | r r r r|}
\hline
   & \multicolumn{4}{|c|}{Arb (real), dot} & \multicolumn{4}{|c|}{Arb (real), approx} \\
$p$ & $\!\!N\!=\!2$ & $\!\!\!\!N\!=\!4$ & $\!\!\!\!N\!=\!8$ & $\!\!\!\!N\!=\!16$ & $\!\!\!N\!=\!2$ & $\!\!\!\!N\!=\!4$ & $\!\!\!\!N\!=\!8$ & $\!\!\!\!N\!=\!16$ \\ \hline
53 & 89 & 65 & 53 & 44 & 64 & 41 & 31 & 23 \\
106 & 98 & 75 & 61 & 53 & 71 & 47 & 37 & 31 \\
212 & 215 & 175 & 159 & 143 & 177 & 147 & 136 & 125 \\
848 & 614 & 571 & 552 & 535 & 567 & 547 & 522 & 507 \\ \hline
\end{tabular}
\label{tab:dotcycles2}
\end{center}
\end{table}


For small $N$, the initialization and finalization overhead
in Algorithm~\ref{alg:dot} is significant.
Table~\ref{tab:dotcycles2} shows that
it nevertheless performs better than a simple loop already for $N = 2$
and quickly converges to the speed measured at $N = 100$.

Algorithm~\ref{alg:dot} does even better with
structured data, for example when the balls are exact, with small-integer coefficients,
or with varying magnitudes.
As an example of the latter, with $N = 1000$, $p = 1024$, and terms $(1/i!) \cdot (\pi^{-i})$,
Algorithm~\ref{alg:dot} takes 0.035 ms while an \texttt{arb\_addmul} loop takes 0.33 ms.

\subsection{Basecase polynomial and matrix operations}

\begin{figure}[tbp]
\begin{center}
\centerline{\includegraphics[width=1.0 \linewidth]{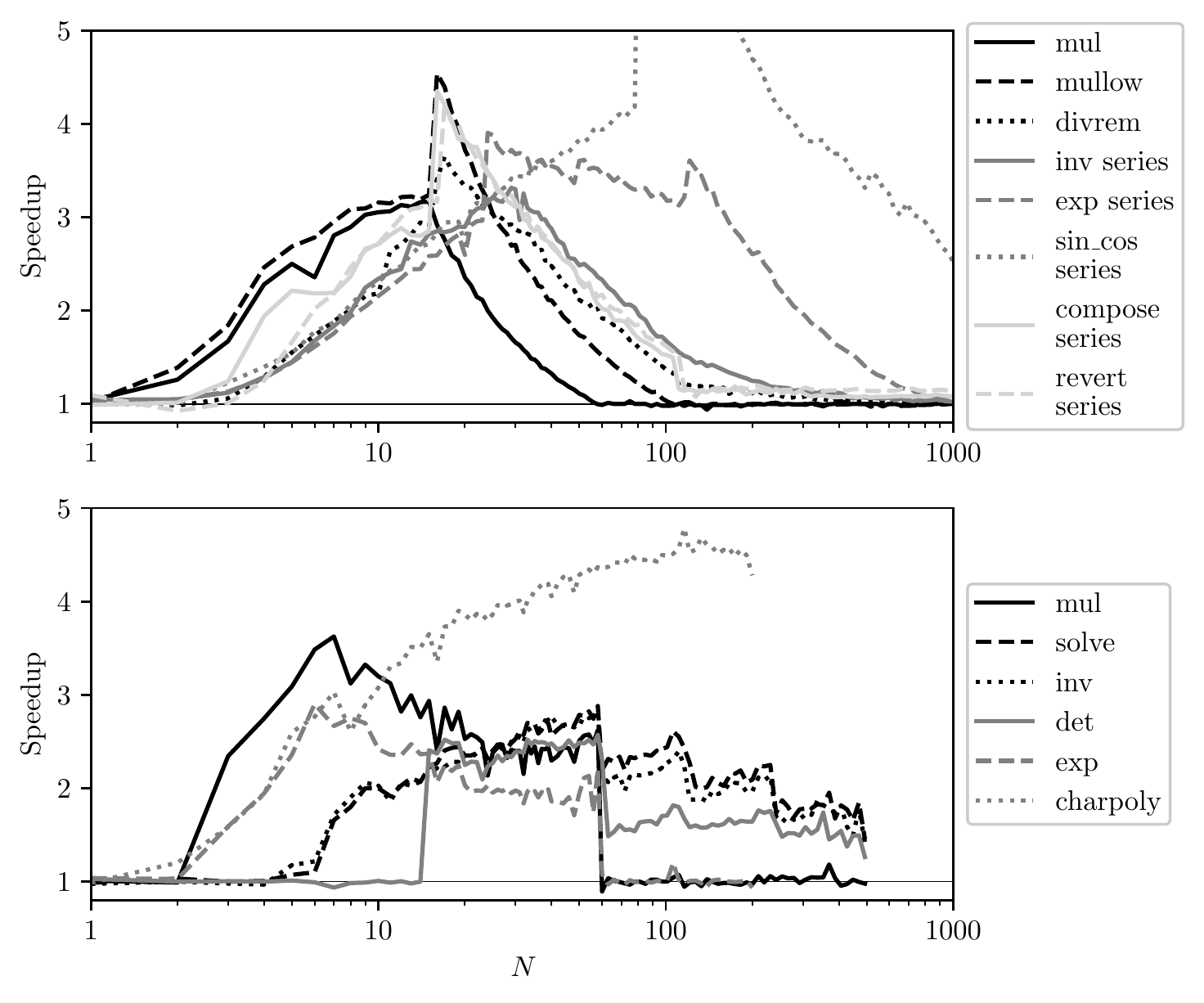}}
\vspace{-2mm}
\caption{Speedup of Arb 2.15 over 2.14 for various operations on
polynomials (top) and matrices (bottom),
here for $p = 64$ and complex coefficients.}
\vspace{-2mm}
\label{fig:bothspeedup}
\end{center}
\end{figure}

The new dot product code was added in Arb 2.15
along with re-tuned cutoffs between small-$N$ and large-$N$ algorithms.
Figure~\ref{fig:bothspeedup} shows the speedup of Arb 2.15 over 2.14
for operations on polynomials and power series of length $N$
and matrices of size $N$, here for $p = 64$ and complex coefficients.

This shows the benefits of Algorithm~\ref{alg:dot},
even in the presence of
a fast large-$N$ algorithm (the block algorithm for matrix multiplication was added in Arb 2.14).
The speedup typically grows with $N$
as the dot product gains an increasing advantage
over a simple multiply-add loop, up to the old cutoff point
for switching to a large-$N$ algorithm.
To the right of this point, the
dot product then gives a diminishing speedup
over the large-$N$ algorithm up to the new cutoff.
Jumps are visible where the old cutoff was suboptimal.
We make some more observations:

\begin{itemize}
\item The speedup around $N \approx$ 10 to 30 is notable since this certainly is a common size for real-world use.
\item Some large-$N$ algorithms like Newton iteration series inversion and block recursive linear solving use recursive operations of smaller size,
so the improved basecase gives an extended ``tail'' speedup into the large-$N$ regime.
\item The power series exponential and sine/cosine improve dramatically.
The large-$N$ method uses Newton iteration which costs several polynomial multiplications,
while the $O(N^2)$ basecase method uses the dot product-friendly
recurrence $\exp(a_1 x + a_2 x^2 + \ldots) = b_0 + b_1 x + b_2 x^2 + \ldots$, $b_0 = 1$, $b_k = (\sum_{j=1}^k (j a_j) b_{k-j}) / k$.
The cutoffs have been increased to
$N = 750$ and $N = 1400$ (for this $p$).
\item The characteristic polynomial (charpoly) does not currently use matrix multiplication in Arb, so we get the pure dot product speedup for large $N$.
\item Series composition and reversion use baby-step giant-steps
methods \cite{BrentKung1978,Johansson2015reversion} where
dot products enter in both
length-$N$ polynomial and size-$\!\sqrt{\!N}$ matrix
multiplications.
\end{itemize}


\subsection{Large-$N$ matrix multiplication}

\begin{table}
\setlength{\tabcolsep}{0.5em}
\caption{Time (s) to multiply size-$N$ matrices. (\#) is the number of blocks $A_s B_s$ used by the block algorithm, where greater than one.}
\begin{center}
\renewcommand{\arraystretch}{1.05}
\begin{tabular}{| r | l l l l | l l | l l l |}
\hline
    &  \multicolumn{4}{c|}{Uniform real} & \multicolumn{2}{c|}{Pascal} & \multicolumn{3}{c|}{Uniform complex} \\ \hline
$p$ &  QD & MPFR\!\! & Arb\!\!\!\!\! & Arb & Arb\!\!\!\!\! & Arb & MPC\!\! & Arb & Arb\\
    &     &      & dot         &  block &  dot   & block & & dot & block\!\!\! \\ \hline
\multicolumn{10}{c}{$N = 100$} \\ \hline
53  &  & 0.035\!\! & 0.019 & 0.0041\! & 0.016 & 0.021 & 0.28 & 0.071\!\! & 0.017\!\! \\
106 &  0.011\!\! & 0.042\!\! & 0.023 & 0.011 & 0.018  & 0.031 & 0.40 & 0.086\!\! & 0.049\!\! \\
212 &  0.11\!\! & 0.11 & 0.061 & 0.021 & 0.063 & 0.046 & 0.50 & 0.23 & 0.092\!\! \\
848 &  & 0.30 & 0.23 & 0.089 & 0.23 & 0.12 & 1.2 & 0.85 & 0.34 \\
\!\!3392 & & 1.7 & 1.7 & 0.48 & 1.7 & 0.55 & 7.1 & 6.1 & 1.9 \\ \hline
\multicolumn{10}{c}{$N = 300$} \\ \hline
53  &  & 0.96 & 0.51 & 0.13 & 0.37 & 0.57 {\scriptsize (3)}\! & 8.1 & 2.0 & 0.37 \\
106 &  0.30 & 1.2 & 0.69 & 0.23 & 0.47 & 0.70 {\scriptsize (3)}\! & 12 & 2.6 & 0.87 \\
212 &  3.0 & 3.0 & 2.2 & 0.34 & 1.8 & 1.2 {\scriptsize (3)}\! & 14 & 7.5 & 1.5 \\
848 &  & 7.9 & 6.2 & 1.2 & 5.1 & 2.4 {\scriptsize (2)}\! & 33 & 26 & 4.9 \\
\!\!3392 &  & 46 & 47 & 6.0 & 44 & 7.3 & 200 & 172 & 24 \\ \hline
\multicolumn{10}{c}{$N = 1000$} \\ \hline
53  &  & 36 & 19 & 3.6 & 12 & 20 {\scriptsize (10)}\! & 313 & 75 & 14 \\
106 &  11 & 44 & 25 & 5.6 & 14 & 23 {\scriptsize (10)}\! & 454 & 97 & 22 \\
212 &  111 & 110 & 76 & 8.2 & 43 & 35 {\scriptsize (9)}\! & 539 & 342 & 33 \\
848 &  & 293 & 258 & 27 & 122 & 80 {\scriptsize (5)}\! & 1230\!\! & 1074\!\! & 107 \\
\!\!3392 &  & 1725\!\! & 1785 & 115 & 1280 & 226 {\scriptsize (2)}\! & 7603\!\! & 6789\!\! & 457 \\ \hline
\end{tabular}
\label{tab:matmul}
\end{center}
\end{table}

Table~\ref{tab:matmul} shows timings to compute $A \cdot A$ where $A$ is a size-$N$ matrix.
We compare two algorithms in Arb (both over balls): \emph{dot} is classical
multiplication using iterated dot products,
and \emph{block} is Algorithm~\ref{alg:block}.
The default matrix multiplication function in Arb 2.16 uses
the \emph{dot} algorithm for $N \le 40$ to $60$ (depending on $p$)
and \emph{block} for larger $N$;
for the sizes of $N$ in the table, \emph{block} is always the default.
We also time QD, MPFR and MPC classical multiplication
(with two basic
optimizations: tiling to improve locality, and
preallocating a temporary inner variable for MPFR and MPC).

We test two kinds of matrices.
The \emph{uniform} $A$ is a matrix where all entries have similar
magnitude. Here, the block algorithm only uses a single block
and has a clear advantage;
at $N = 1000$, it is
5.3 times as fast as the classical algorithm when $p = 53$
and 16 times as fast when $p = 3392$.

The \emph{Pascal} matrix $A$ has entries
$\pi \cdot {i+j \choose i}$ which vary
in magnitude between unity and $4^N$.
This is a bad case for Algorithm~\ref{alg:block}, requiring many blocks when $N$ is
much larger than $p$.
Conversely, the classical algorithm is faster for this matrix than for the uniform matrix
since Algorithm~\ref{alg:dot} can discard many input limbs.
In fact, for $p \le 128$ the classical algorithm is roughly 1.5 times
as fast as the block algorithm
for $N$ where Arb uses the block algorithm by default, so the
default cutoffs are not optimal in this case.
At higher precision, the block algorithm does recover the
advantage.

\subsection{Linear solving, inverse and determinants}

\begin{table}
\caption{Time (s) to solve a size-$N$ real linear system in arbitrary-precision arithmetic.
* indicates that the slower but more accurate Hansen-Smith algorithm is used.}
\begin{center}
\renewcommand{\arraystretch}{1.05}
\begin{tabular}{|r r | l l | l l |}
\hline
$N$ & $p$ & Eigen & Julia & Arb (approx)\!\!\!\! & Arb (ball) \\ \hline
10 & 53 & 0.00028 & 0.000066 & 0.000021 & 0.00013* \\
10 & 106 & 0.00029 & 0.000070 & 0.000025 & 0.000040 \\
10 & 212 & 0.00033 & 0.00010 & 0.000055 & 0.000074 \\
10 & 848 & 0.00043 & 0.00022 & 0.00014 & 0.00016 \\
10 & 3392 & 0.0012 & 0.0010 & 0.00088 & 0.00090 \\ \hline
100 & 53 & 0.051 & 0.064 & 0.0069 & 0.040* \\
100 & 106 & 0.054 & 0.070 & 0.0084 & 0.049* \\
100 & 212 & 0.080 & 0.10 & 0.024 & 0.10* \\
100 & 848 & 0.16 & 0.22 & 0.080 & 0.35* \\
100 & 3392 & 0.71 & 0.90 & 0.49 & 0.50 \\ \hline
1000 & 53 & 37 & 301 & 2.3 & 13* \\
1000 & 106 & 39 & 401 & 3.3 & 20* \\
1000 & 212 & 64 & 488 & 6.6 & 36* \\
1000 & 848 & 132 & 947 & 24 & 118* \\
1000 & 3392 & 601 & 2721 & 153 & 609* \\ \hline
\end{tabular}
\label{tab:solve}
\end{center}
\end{table}

Arb contains both approximate floating-point and ball
versions of real and complex triangular solving,
LU factorization, linear solving and matrix inversion.
All algorithms are block recursive, reducing the work to matrix multiplication
asymptotically for large $N$ and to dot products (in the form of basecase triangular solving
and matrix multiplication) for small $N$.
Iterative
Gaussian elimination is used for $N \le 7$.

In ball (or interval) arithmetic, LU factorization
is unstable and
generically loses $O(N)$ digits even for a well-conditioned matrix.
This problem can be fixed with preconditioning~\cite{rump2010verification}.
The classical Hansen-Smith algorithm \cite{hansen1967interval}
solves $AX = B$ by first computing an approximate
inverse $R \approx A^{-1}$ in floating-point arithmetic
and then solving $(R A) X = R B$ in interval or ball arithmetic.
Direct LU-based solving in ball arithmetic behaves nicely
for the preconditioned matrix $R A \approx I$.

Arb provides three methods for linear solving in ball arithmetic:
the LU algorithm, the Hansen-Smith algorithm,
and a default method using LU when $N \le 4$ or $p > 10N$
and Hansen-Smith otherwise.
In practice, Hansen-Smith is typically 3-6 times as slow as
the LU algorithm.
The default method thus attempts to give good performance both for
well-conditioned problems (where low precision should be sufficient)
and for ill-conditioned problems (where high precision is required).
Similarly, Arb computes determinants using ball LU factorization
for $N \le 10$ or $p > 10 N$
and otherwise via preconditioning using approximate LU factors~\cite{rump2010verification}.

Table~\ref{tab:solve} compares speed for solving $AX=B$ with a
uniform well-conditioned $A$ and a vector $B$.
Due to the new dot product and matrix
multiplication,
the LU-based approximate solving in Arb is significantly faster than
LU-based solving with MPFR entries
in both the Eigen 3.3.7 C++ library~\cite{guennebaud2018} and Julia 1.0 ~\cite{bezanson2017julia}.
The verified ball solving in Arb is also competitive.
Julia is extra slow for large $N$ due to garbage collection, which
incidentally makes an even bigger case for an atomic dot product avoiding
temporary operands.

\subsection{Eigenvalues and eigenvectors}

\begin{table}[tbp]
\caption{Time (s) for eigendecomposition of size-$N$ complex matrix}
\begin{center}
\renewcommand{\arraystretch}{1.05}
\begin{tabular}{|r r | l | l l l|}
\hline
$N$ & $p$ & Julia & Arb (approx)\!\!\! & Arb (Rump)\!\! & Arb (vdHM)\! \\ \hline
10 & 128 & 0.021 & 0.0036 & 0.0082 & 0.0045 \\
10 & 384 & 0.043 & 0.011 & 0.022 & 0.013 \\
100 & 128 & 8.8 & 2.5 & 18.2 & 2.9 \\
100 & 384 & 18.5 & 8.7 & 59 & 9.8 \\
1000 & 128 & $>3\!\cdot\!10^4$ & 2764 & & 2981 \\
1000 & 384 &                   & 9358 & & 9877 \\ \hline
\end{tabular}
\label{tab:eig}
\end{center}
\end{table}

Table~\ref{tab:eig} shows timings for computing 
the eigendecomposition of the matrix with entries $e^{i (jN+k)^2}, 0 \le j, k < N$.
Three methods available in Arb 2.16 are compared.
The \emph{approx} method is
the standard QR algorithm~\cite{kressner2005} (without error bounds),
with $O(N^3)$ complexity.
We include as a point of reference timings
for the QR implementation in the Julia package GenericLinearAlgebra.jl
using MPFR arithmetic.
The other two Arb methods compute rigorous enclosures in ball arithmetic
by first finding
an approximate eigendecomposition using the QR algorithm
and then performing a verification using ball matrix multiplications and linear solving.
The \emph{Rump} method~\cite{rump2010verification} verifies one
eigenpair at a time requiring $O(N^4)$ total operations, and
the \emph{vdHM} method~\cite{vdH:ball,van2017efficient} verifies
all eigenpairs simultaneously in $O(N^3)$ operations.

The kernel operations in the QR algorithm are rotations
$(x, y) \gets (c x + s y, \, \overline{c} y - \overline{s} x)$,
i.e.\
dot products of length~2, which we have only improved slightly in this work.
A useful future project would be an arbitrary-precision
QR implementation with block updates to
exploit matrix multiplication.
Our work does already speed up the initial reduction to Hessenberg form
in the QR algorthm, and it
speeds up both verification algorithms;
we see that the \emph{vdHM} method
only costs a fraction more than the unverified \emph{approx} method.
The \emph{Rump} method is more expensive but gives more precise balls than \emph{vdHM};
this can be a good tradeoff in some applications.

\section{Conclusion and perspectives}

We have demonstrated that optimizing the dot product as an atomic
operation leads to a significant reduction in overhead for
arbitrary-precision arithmetic, immediately speeding up
polynomial and matrix algorithms.
The performance is competitive with non-vectorized double-double and
quad-double arithmetic, without the drawbacks of these types.
For accurate large-$N$ matrix multiplication, using scaled integer blocks
(in similar fashion to previous work for polynomial multiplication)
achieves even better performance.

It should be possible to treat
the Horner scheme for polynomial evaluation in similar way
to the dot product, with similar speedup.
(The dot product is itself useful for polynomial evaluation,
in situations where powers of the argument can be recycled.)
More modest improvements should be possible
for single arithmetic operations in Arb.
See also \cite{van2016evaluating}.

In addition to the ideas for algorithmic improvements already noted in this paper,
we point out
that Arb would benefit from faster integer matrix multiplication
in FLINT.
More than a factor two can be gained with better residue conversion code
and use of BLAS~\cite{doliskani2018simultaneous,giorgi2014toward}.
BLAS could also be used for the radius matrix multiplications in Arb
(we currently use simple C code since the FLINT multiplications
are the bottleneck).

The FLINT matrix code is currently single-threaded,
and because of this, we only benchmark single-core performance. 
Arb does have a multithreaded version of classical matrix multiplication
performing dot products in parallel, but this code is typically not
useful due to the superior single-core efficiency
of the block algorithm. Parallelizing
the block algorithm optimally is of course the more interesting problem.


\bibliographystyle{plain}
\bibliography{references}

\end{document}